\documentclass[twoside]{report}
\usepackage{iwsm}
\usepackage{graphicx}
\usepackage{amsmath, amssymb}
\usepackage{booktabs}

\begin{document}


\title{Resolving the Lord's Paradox}
\titlerunning{Lord's Paradox}

\author{Priyantha Wijayatunga\inst{1}}
\authorrunning{Wijayatunga et al.}    

\institute{Department of Statistics,  Ume\r{a} University,  Ume\r{a} SE-90187, Sweden}

\email{priyantha.wijayatunga@umu.se}

\abstract{An explanation to Lord's paradox using ordinary least square regression models is given. It is not a paradox at all, if the regression parameters are interpreted as predictive or as causal with stricter conditions and be aware of laws of averages.  We use derivation of a super-model from a given sub-model, when its residuals can be modelled with other potential predictors as a solution.}

\keywords{Effect; Predictive; Causal; Confounding.}

\maketitle



\section{Introduction}
In 1967 Frederic Lord posed following question (see Lord 1967 and  Pearl 2016) that became a paradox among applied statistical community.  To see effects and if there is any sex difference of diet provided in a university weights of students at time of their arrival and those a year later are recorded. The data are independently examined by two statisticians. The first examines the mean weight of the girls at the beginning and at the end of the year, and finds that they are to be identical, i.e.,  frequency distribution of the weight for the girls is not changed, so is for the boys. The second statistician finds that the slope of the regression line of the final weight on the initial weight is essentially the same for both sexes but the regression coefficient of the variable sex to be statistically significant and concludes that the boys showed significantly more gain in weight than the girls when proper allowance is made for differences for initial weight. 

Conclusions of the two statisticians seem to contradict with each other; the first is predictive and the second is both predictive, and causal if the initial weight is the only confounder of causal relation between the sex and the final weight. The second has given causal effect of the sex on the final weight (weight gain) by a regression coefficient. In fact, to give it by comparing, two supports of the confounder of both sexes should coincide.  But one can assume that the population initial weight ranges of boys and girls coincide even though sample counterparts differ (so, extrapolation is meaningful).   

Let the initial weight, final weight and sex are denoted by $W_I$, $W_F$, $S$ respectively ($S=0,$ a girl and $S=1,$ a boy) and weight gain  be $D=W_F - W_I.$   If the effect of $S$ on $D$ is found by difference of conditional means, $E\{D\vert S=1\}-E\{D \vert S=0\}$ then it is no effect. This can be found by running regression of $D$ on $S.$  Note that $E\{ W_F \vert S=1\} =E\{ W_I \vert S=1\}$, (say,  $\mu_B$) and $E\{ W_F \vert S=0\} =E\{ W_I \vert S=0\}$, (say, $\mu_G$).  If $E\{D \vert W_I=i,S=1\}$ and $E\{D \vert W_I=i,S=0\}$  are calculated simply by partitioning the data by  taking $W_I$ to be discrete or as a functions of $i$, then the difference $E\{D \vert W_I=i,S=1\}-E\{D \vert W_I=i,S=0\}$ may not be zero for each $i$, so may be difference of their weighted  means, $\sum_i E\{D \vert W_I=i,S=1\}p(W_I=i)-\sum_i E\{D \vert W_I=i,S=0\}p(W_I=i).$  If the effect of $S$ on $W_F$ is calculated by it then it is different from former value (paradoxical!). 

Now let us see why two types of differences of averages differ by simple algebra, that will say that they should have two different  interpretations. First assume that we have $a$ number of subgroups of boys and, for simplicity, the same is true for girls. Let $D_{ij}^1$ be the weight gain of the $j$-th boy in the $i$-th subgroup of boys where sub-group size is $n_i$ and $D_{ij}^0$ be that of the girls where sub-group size is $m_i$ and furthermore, let $f_i^1=n_i / \sum_k n_k$,  $f_i^0=m_i / \sum_k m_k$ and  $f_i =(n_i+m_i) / \sum_k (n_k+m_k)$  for $j=1,...,n_i $ and $ i=1,...,a.$ And let $A_1$ be difference of the average weight gain of the boys and the girls, $\bar{D}_i^1= \sum_j D_{ij}^1 /n_i$ and $\bar{D}_i^0= \sum_j D_{ij}^0 /m_i$ for $i=1,...,a.$ So, 
\begin{eqnarray*}
A_1 &=& \frac{ \sum_{i=1}^a \sum_{j=1}^{n_i} D_{ij}^1}{\sum_{i=1}^a n_i}  - \frac{\sum_{i=1}^a \sum_{j=1}^{m_i} D_{ij}^0}{\sum_{i=1}^a m_i} 
= \sum_{i=1}^a \bar{D}_i^1 f_i^1 - \sum_{i=1}^a  \bar{D}_i^0 f_i^0\\
   &\neq & \frac{1}{2} \Big\{\sum_{i=1}^a \bar{D}_i^1 f_i^1 + \sum_{i=1}^a \bar{D}_i^1 f_i^0  - \sum_{i=1}^a  \bar{D}_i^0 f_i^0 - \sum_{i=1}^a \bar{D}_i^0 f_i^1 \Big\}; \textrm{    generally }\\
   &=& \frac{1}{2} \Big\{  \sum_{i=1}^a (\bar{D}_i^1 - \bar{D}_i^0)( f_i^1+f_i^0)  \Big\} \neq  \sum_{i=1}^a (\bar{D}_i^1 - \bar{D}_i^0) f_i =A_2
\end{eqnarray*}
where $f_i= \alpha f_i^1+ (1-\alpha) f_i^0$ for $i=1,...,a$ such that $\alpha=\sum_{i=1}^a n_i /\sum_{i=1}^a (n_i+m_i)$ and $A_2$ is the difference of weighted averages of the sub-group weight gain averages. So, the difference of group averages $A_1$ (which is zero in our case) is  different from the difference of  pooled-weighted average of the  sub-group averages $A_2$.  The second statistician compares the boys and the girls subgroup-wise and finds that it is a constant gain for the boys over the girls across the subgroups, i.e., $\bar{D}_i^1 - \bar{D}_i^0$ is constant for all initial weight $i.$  Therefore he finds that the boys gain more weight than  the girls in corresponding sub-groups. Note that for simplicity we have taken initial weights as discrete values. In fact, $ A_2 = \sum_i E\{D \vert S=1, W_I=i\}p(W_I=i)-\sum_ i E\{D \vert S=0, W_I=i\}p(W_I=i)  $ is the causal effect of $S$ on $D$ if $W_I$ is the only confounder, under the linear assumption. It is different from $A_1$ unless $E\{D \vert S, W_I\}=E\{D \vert S \}.$ The confounding effect ($A_2-A_1$) depends on how different $f^1$ and $f^0$ are (can have a measure from them).

\section{Regression Solution}

Now we define interpretation of ordinary least square (OLS) estimates of the regression coefficients (parameters).  The OLS estimation  is based on the variation of the response variable $Y$  for a given functional form of the values of explanatory factors.  Regression coefficients are estimated so that sum of squared prediction errors for the data in the sample is the minimum. So,  reverse regression is not generally obtainable from forward regression and may not be consistent with the latter.   For simple linear regression one can easily establish that the reverse regression and the forward regression are consistent with each other if and only if one of the regressions have symmetric residuals  about and uni-modal at conditional expectation of response, that implies other regression too. 

Now consider the OLS linear regression model $Y=\beta_0+\beta_1 X_1+\beta_2 X_2+ \epsilon $, then linear effect of $X_1$ on $Y$ when $X_2$ is held unchanged  is given by  $\beta_1$ if $Y$ values are symmetric about and uni-modal at $\beta_0+\beta_1 X_1+\beta_2 X_2$. It is clear that the supports of $X_2$ for each value of $X_1$ are the same (or extrapolation is meaningful if empirical supports  differ). Symmetry and uni-modality of $Y$ values for given values of $X_1$ and $X_2$ are observed if all other factors that affect or are associated, but are not taken into consideration are allowed to vary pure randomly. This is a fundamental assumption used in statistical modelling often implicitly.  
 
Let us do a regression of $W_F$ on the binary variable $S$. Then we get the model $W_F = \mu_G +  (\mu_B - \mu_G) S + \epsilon_1. $ where the regression co-efficient of $S$ is the predictive effect of $S$ on $W_F$  provided that above requirement is fulfilled. The residuals of the model are just individual values of $D$, i.e., $\epsilon_1=D$ for each subject and it is easy to see in Fig. 1 of Lord 1967, that the residuals are predictive by $W_I$ for each sex category separately, $\epsilon_1 \not\perp W_I \vert S.$  However, it may be that $\epsilon_1 \perp  S.$ So, if the two clusters of values of $W_F$ for two sexes  are symmetric about and uni-modal at the respective means then the effect of $S$ on $W_F$ is the regression coefficient of $S$ in the model. But it is uncontrolled confounders that are associated with $W_F$, then it should be interpreted accordingly.  That is, it is the predictive effect of sex differences and causal if there are no confounders such as $W_I$. And we see that we get zero predictive effect from the meal change since the regression coefficient is the same as that when the girls and boys had previous meal type. 

Let we can write the distribution of residuals for each value $s$ of $S,$ say, $f(\epsilon_1 \vert s)$ as a mixture, 
$ f(\epsilon_1 \vert s) =\int g(\epsilon_1 \vert x,s) \pi(x,s) dx $
for some random variable $X$, and for each value $x$ of $X$ the component distribution $g(\epsilon_1 \vert x,s)$ may have non-zero mean such that 
$E\{\epsilon_1 \vert s \} =\int E\{ \epsilon_1 \vert x,s\}\pi(x,s) dx=0$
and then we have that 
$Var\{\epsilon_1 \vert x,s\} \leq Var\{\epsilon_1 \vert s \}$
where $\pi(x,s) = h(x \vert s)p(s);$ here $h(x \vert s)$ is the conditional probability density of $X$ given $S=s$ and $p(s)$ is the marginal probability distribution of $S.$ If $X$ could be identified meaningfully, then model should include such feature variables too. In this case,  $X$ could be identified as the  initial weight $W_I$.  Then one should accept the upgraded  model that includes  $W_I$ too. It has  residuals that have a smaller conditional standard deviation given $W_I$ and $S.$  Furthermore, if $W_I$ is the only confounding factor and when it is also included in the model the  the coefficient of $S$ is the causal effect of $S$ on $W_F$.

Let the residual $\epsilon'_1$  corresponds to the context that $W_I=w_I$ and $S=s$ and then it can be written as $\epsilon'_1=\mu_{\epsilon_1}^{w_I,s} + \epsilon_2$ where $\mu_{\epsilon_1}^{w_I,s}$ is the expectation of it. So, we have $E\{\epsilon_2 \vert W_I=w_I,S=s\}=0$ and also that $Var\{\epsilon_2 \vert W_I=w_I,S=s\} \leq Var\{\epsilon_1 \vert S=s\}.$   And furthermore, we can have that $\mu_{\epsilon_1}^{w_I,0}=a_0+b_0 w_I$ for $s=0$ and $\mu_{\epsilon_1}^{w_I,1}=a_1+b_0 w_I$ for $s=1$ where $a_0,b_0$ and $a_1$ are constants.  Now, given that $W_I=w_I$ and $S=s$, for $s=0,1$, and  $I(A)=1$ when $A$ is a true statement and $I(A)=0$ otherwise, we have
\begin{eqnarray*}
 W_F &=& W_F = \mu_G +  (\mu_B - \mu_G) s + \mu_{\epsilon_1}^{w_I,s} + \epsilon_2 \\
   &=& \mu_G +  (\mu_B - \mu_G) s +(a_0+b_0 w_I)I(S=0)+ (a_1+b_0 w_I)I(S=1) + \epsilon_2\\
         &=&  \mu_G +  (\mu_B - \mu_G) s +a_0 I(S=0)+ a_1 I(S=1) + b_0 w_I+ \epsilon_2\\
         &=&  \mu_G +  (\mu_B - \mu_G) s  +a_0 (1-s)+ a_1 s + b_0 w_I+ \epsilon_2\\
         &=&  \mu_G +  a_0+ (\mu_B - \mu_G - a_0 + a_1) s + b_0 w_I+ \epsilon_2
\end{eqnarray*} 
So we can obtain a super-model (regression) from a given regression model (it is a sub-model of the former) as long as its residuals are predictive (linearly in this case) with another explanatory variable. The predictive effect of $S$ on $W_F$ when controlled for $W_I$ is $\mu_B - \mu_G - a_0 + a_1$ that is generally different from earlier value of $\mu_B -\mu_G$ and for each individual model prediction is  more accurate than that of the previous model, therefore new model is preferred to the previous one. If $W_I$ is only a confounder but not an intermediate variable between the causal pathway between $S$ and $W_F$, and has a common support for all values of $S$, then $\beta_1$ is the average causal effect of $S$ on $W_F$ in the linear case. In our example, sample supports of $W_I$ for $S=1$ and $S=0$ differ but we can assume that they are the same in the population (so, extrapolation is meaningful). Note that the above arguments can be generalised. For restrictions of space, we avoid presenting solution to the paradox, that is based on causal diagrams. We object recent solution by Pearl. Our explanations  comply with Lord's initial comments.




\references
\begin{description}

\item[Lord, F. M.](1967). \label{Lord:paradox}
A Paradox in the Interpretation of Group Comparisons. 
{\it Psychological Bulletin}, {\bf 68}(5), 304\,--\,305.

\item[Pearl, J. ](2016). \label{Pearl:causal}
Lord's Paradox Revisted - (Oh Lord Kumbaya!). 
{\it Journal of Causal Inference}, {\bf 4}(2). DOI: 10.1515/jci-2016-0021.
\end{description}

\end{document}